\documentclass[12pt]{article}
\usepackage{amssymb}
\usepackage{graphicx}

\newtheorem{proposition}{Proposition}
\newtheorem{theorem}{Theorem}

\begin{document}

\begin{center}

{\Large\bf New Applications of Quantum Algebraically\\
 Integrable Systems in Fluid Dynamics} \\

\vspace{5mm}

{ Anne Boutet de Monvel, Igor Loutsenko, Oksana Yermolayeva}\\

\vspace{5mm}

Institut de Math\'ematiques de Jussieu, UPD Paris VII\\
Institut Henri Poincar\'e, UPMC Paris VI\\

\vspace{5mm}

e-mail: aboutet@math.jussieu.fr, loutsenko@math.jussieu.fr, yermolay@ihp.jussieu.fr

\vspace{5mm}

Abstract\\

\end{center}

\begin{quote}
The rational quantum algebraically integrable systems are non-trivial
generalizations
 of Laplacian operators to the case of elliptic operators with variable coefficients.
 We study corresponding extensions of Laplacian growth connected with algebraically
 integrable systems, describing viscous free-boundary flows in non-homogenous media.
 We introduce a class of planar flows related with application of Adler-Moser polynomials and
 construct solutions for higher-dimensional cases, where the conformal mapping technique is unavailable.
\end{quote}

\begin{section}
{ Introduction}

We start with an inverse problem in fluid dynamics.
Before posing the problem, we explain general settings and notations.

Consider a viscous flow or a flow of liquid in a thin (not necessarily planar) layer of
non-homogeneous porous medium. The layer can be viewed as a
2-dimensional surface embedded in the 3-dimensional Euclidean
space. We let the layer curvature, permeability, porosity and
thickness depend on the surface spatial coordinates $x,y$. We can
choose $x,y$ such that locally
$$
dl^2=G(x,y)(dx^2+dy^2)
$$
where $dl$ is the surface length element. The surface area element
is $d\Sigma=G dxdy$, and the volume of the liquid that can be
absorbed in the range $x+dx,y+dy$ equals
$$
dV=\eta h d\Sigma=\eta hGdxdy,
$$
where $\eta=\eta(x,y), h=h(x,y)$ are the medium porosity and the
layer thickness, respectively.

In the porous medium, the flow velocity $v=(dx/dt,dy/dt)$ is
proportional to the gradient $\nabla=(\partial/\partial x,
\partial/\partial y)$ of the pressure $P$
$$
v=-\frac{\kappa}{\sqrt{G}}\nabla P,
$$
where $\kappa=\kappa(x,y)$ is the medium permeability.

It is seen from the above that only the two combinations of variable
coefficients, namely
$$
\eta h G, \quad \frac{\kappa}{\sqrt{G}}
$$
enter the equation of flow motion, and it is convenient to absorb
$h$ and $G$ into definitions of the other coefficients. Therefore,
without loss of generality, we can consider flow in the plane
parametrized by the complex coordinates $z=x+iy, \bar z=x-iy$.
For this purpose we choose $\eta$ and $\kappa$ to depend on $z, \bar z$, while setting the
remaining coefficients to unity. It is also convenient to redefine the pressure by $P\to -P$.

The liquid volume conservation
leads to the continuity equation
\begin{equation}
(\nabla\cdot\eta v)=0 \label{continuity},
\end{equation}
while the dynamic law of motion rewrites as
\begin{equation}
v=\kappa\nabla P \label{darcy}.
\end{equation}
We consider the situation where the liquid occupies a bounded,
simply-connected open region $\Omega$ of the plane, whose time
evolution $\Omega=\Omega(t)$ is induced by the flow.

At fixed time $t$, the pressure is constant along the boundary
\begin{equation}
P(\partial\Omega(t))=P_0(t).
\label{boundary}
\end{equation}
Note, that the in the case of simply-connected domains
considered here, the dynamics is independent of $P_0(t)$, and for
convenience the latter can be set to zero.

The normal velocity of the boundary $v_n$ and that of the flow
coincide at $\partial\Omega$
\begin{equation}
v_n=n\cdot v\quad {\rm if} \quad z\in
\partial\Omega \label{kinematik}.
\end{equation}
The flow is singularity driven. For instance
\begin{equation}
P\to \frac{q(t)}{\kappa(z_1,\bar z_1)\eta(z_1,\bar
z_1)}\log|z-z_1|+\sum_{j=1}^
k\left(\frac{\mu_j(t)}{(z-z_1)^j}+\frac{\bar\mu_j(t)}{(\bar z-\bar
z_1)^j}\right), \quad {\rm as} \quad z \to z_1,
\label{sors}
\end{equation}
when $k+1$ multipole sources are located in the vicinity of $z=z_1$ in the interior of the fluid domain $\Omega$.

Equations (\ref{continuity}) - (\ref{sors}) define the free
boundary problem where evolution of the boundary $\partial
\Omega(t)$ is completely determined by the initial condition
$\partial \Omega(0)$ and the dynamics (or strengths) $q(t)=\bar q(t), \mu_j(t),
 j=1,..,k$ of the sources.

It follows from equations (\ref{continuity}), (\ref{darcy}) and (\ref{sors}) that the pressure satisfies the following linear PDE
\begin{equation}
(\nabla \kappa\eta \nabla) P = 2\pi\hat q [ \delta(x-x_1) \delta(y-y_1) ]
\label{Pequation}
\end{equation}
where $\hat q$ is a differential operator of the order $k$
\begin{equation}
\hat q=q(t)+\sum_{j=1}^k (-1)^j\left(
q_j(t)\frac{\partial^j}{\partial z^j}+\bar
q_j(t)\frac{\partial^j}{\partial \bar z^j}\right)
\label{q}
\end{equation}
and $\delta$ denotes Dirac delta-function.

Since the permeability and porosity are always positive, the operator in the lhs of the last equation is elliptic and this class of free boundary problems is called elliptic growth. For the case of homogenous media one has
\begin{equation}
\kappa={\rm const}, \quad \eta={\rm const},
\label{homogenous}
\end{equation}
the above operator is Laplacian and a subclass of related growth processes is called Laplacian growth or Hele-Shaw problem (see e.g. \cite{VE}).

\vspace{0.5cm}

{\bf Inverse problem}: Consider an inhomogeneous medium with given
permeability $\kappa(x)$ and porosity $\eta(x)$, depending on one
coordinate only. This is the typical case for many applications
corresponding, e.g. to filtration flows in stratified media. In this
problem one wishes to control the boundary dynamics using a finite
number of closely placed fixed sources/sinks by varying their
strengths in time.

Our approach to this problem is to find exact solutions that better
approximate the dynamics in quest. Such exact dynamics can be
constructed for a wide class of $\kappa(x)$ and $\eta(x)$ related to
rational solutions of the KdV hierarchy.

The shape of the domain is approximated by a time dependent polynomial
conformal mapping
\begin{equation}
z=F(w,t)=z_1+r(t)w+\sum_{i=2}^l u_i(t)w^i, \quad \bar r=r,
\label{conformal}
\end{equation}
from the unit circle $|w|=1$ in the "mathematical" $w$-plane.

Such  a polynomial dynamics  can be sustained by a finite number of
closely placed sources and sinks when the permeability and porosity
are equal to:
\begin{equation}
\kappa \eta = \xi(x)^{-2}, \quad \xi(x) =\frac {p_n(x)}{p_{n-1}(x)},
\quad \eta(x) = p_{n-1}(x) S(x).
\label{etakappa}
\end{equation}
Here $p_n$ is the $n$th Adler-Moser polynomial and $S(x)$
is a polynomial with free coefficients that approximates
characteristics of the medium inside the domain $\Omega$.

The number of singularities (sources/sinks) depends both on the degree
of conformal mapping and the properties of the medium.
\begin{equation}
k+1=\left( \deg(S(x))+\frac{n(n+1)}{2}+1\right) \deg(F(w)).
\label{sources}
\end{equation}

The Adler-Moser polynomials were considered for the first time by
Burchnall and Chaundy  \cite{BC} as solutions of the bilinear
differential recurrence relation \footnote{They were later
rediscovered by Adler and Moser as rational solutions of KdV
hierarchy \cite{AM}}
\begin{equation}
p_n(x) p_{n+1}''(x)-2p_n'(x)p_{n+1}'(x)+p_{n+1}(x) p_n''(x)=0, \quad
p_0=1
\label{BC}
\end{equation}
At each step of this recurrence procedure a new nontrivial integration
constant is introduced \footnote{The second integration constant is the common scaling factor which, without loss of generality, can be set to unity}  and therefore, the intrinsic property of
Adler-Moser polynomials $p_n(x)$ consists in possessing $n+1$ real
free parameters. Since two of them are coming from shift and
scaling, the polynomials are effectively defined by the $n-1$ free
parameters. This, together with the freedom in definition of $S(x)$,
allows an approximation of the stratified medium characteristics.
An increase in the quality of approximation will, in general, result in
increase of the number of sources, as seen from (\ref{sources}).

The above properties make the class of elliptic growth related to
the Adler-Moser polynomials  more important for applications than
that studied by us in connection with integrable
systems related to finite reflection groups \cite{L}.

\vspace{0.5cm}

{\bf Example:} Let us consider the simplest example with the linear porosity $\eta(x)=x$. In this case the first two Adler-Moser polynomials
$$
p_1 = x,  \quad p_2 = x^3 +t_1
$$
can approximate the dynamics with $\kappa(x)$ being
 $$
 \kappa = \frac{x}{(x^3+t_1)^2}
 $$
 Then, according to (\ref{sources}),
 $$
 k+1= 4 \deg(z(w)).
 $$
For instance, 4 sources are needed to sustain the circular shape of
the domain, i.e. $z=z_1+r(t)w$, for $\kappa(x)$ of the present example.

\vspace{0.5cm}


In the sequel we will describe important connections between the
theory of quantum algebraically integrable systems (QAIS),
integrable elliptic growth and generalized quadrature identities.
Later, we will consider the higher-dimensional cases, when the conformal
mapping technique is inapplicable and the result is uniquely due to
advances in the theory of QAIS.

\end{section}

\begin{section}
{ Quantum algebraically integrable systems\\ as generalisations of
Laplace operators}

Let us review some common facts on QAIS. The classical Liouville integrability is equivalent to the existence of a maximal set of Poisson commuting invariants in a Hamiltonian system: For a given classical integrable system  with $n$ degrees of freedom there cannot exist more than $n$ functionally independent integrals in involution.

In the case of quantum integrable system in $n$ dimensions,
integrals form commutative rings of differential operators
generated by $n$ independent elements and such a ring is called
complete. Algebraic quantum integrability implies the existence of additional,  independent generators of the ring of
commuting differential operators. Such rings of commuting
differential operators are called overcomplete.

Note, that results of the present article hold  for
algebraically integrable differential operators with coefficients
that are rational functions of coordinates. So, under algebraically
integrable systems we mean here only rational QAIS \footnote{The most general class of algebraically integrable operators is represented by operators having elliptic function coefficients, for which trigonometric and rational coefficients appear in limiting cases. They are also called the algebro-geometric operators \cite{C}.}.

The simplest example of nontrivial QAIS comes from the theory of
Calogero-Moser operators in one dimension. Consider a quantum
system with the Hamiltonian
$$
H = \partial_x^2 - \frac{2}{x^2}.
$$
For our purposes, it is convenient to consider a gauge-equivalent
differential operator
$$
L_1=xHx^{-1}=x^2\partial_x\frac{1}{x^2}\partial_x=\partial_x^2-\frac{2}{x}\partial_x
$$
For this system, there exists another differential operator
$$
M_1 = \partial_x^3 - \frac{3}{x} \partial_x^2 + \frac{3}{x^2}
\partial_x
$$
that commutes with $L_1$ and does not belong to the ring generated by $L_1$. Thus, in one dimension,
we have an overcomplete commutative ring generated by two independent elements $L_1$ and $M_1$. It is important to note that this ring is not trivial in the sense that it is not a subring of any ring of commuting differential operators generated by a single element, since the above system does not possess a first order integral.

The following example concerns 2-dimensional case. In $d=2$ the ring may be formed by the following set of commuting differential operators:
\begin{equation}
L_1 =x^2\nabla\frac{1}{x^2}\nabla =
x^2\partial_x\frac{1}{x^2}\partial_x+\partial_y^2, \quad M_1 =
\partial_x^3 - \frac{3}{x} \partial_x^2 + \frac{3}{x^2}
\partial_x, \quad K_1
=
\partial_y^2. \label{Lone}
\end{equation}
Let us note, that the operator $L_1$ is not trivial in the sense that it is not
equivalent, up to a change of dependent and independent variables and multiplication by a function, to any Beltrami-Laplace operator on some 2-dimensional manifold. The operator $L_1$ describes elliptic growth with $\kappa\eta=1/x^2$ in (\ref{Pequation}).

Another important concept of algebraic integrability is a notion of an intertwining operator. Consider the differential identity
\begin{equation}
 T \Delta = L T
\label{ii}
\end{equation}
that relates the Laplace operator with a second order differential operator $L$. The operator $T$ is called an intertwining operator and (\ref{ii}) is an intertwining identity.

The simplest example of an intertwining identity is related to factorization of the one-dimensional Laplace operator $\partial_x^2 \quad$:
$$
\frac{\partial^2}{\partial
x^2}=\left(\frac{1}{x}\frac{\partial}{\partial
x}\right)\left(x\frac{\partial}{\partial x}-1\right).
$$
and
$$
\left(x\frac{\partial}{\partial x}-1\right){\partial_x^2}=\nonumber \left(x\frac{\partial}{\partial
x}-1\right)\left(\frac{1}{x}\frac{\partial}{\partial
x}\right)\left(x\frac{\partial}{\partial x}-1\right)
$$
from which we obtain the simplest intertwining identity
\begin{equation}
T_1\Delta=L_1T_1 \quad \quad  T_1=x\frac{\partial}{\partial x}-1
\label{inter_1}
\end{equation}
that relates  Laplacian with the Calogero-Moser operator $L_1$ from (\ref{Lone}).

The higher intertwining identities
$$
T_n\Delta=L_nT_n
$$
involve operators of higher order, like in the following example:
\begin{equation}
T_n=x^n\left(\frac{\partial}{\partial
x}-\frac{n}{x}\right)\left(\frac{\partial}{\partial
x}-\frac{n-1}{x}\right)...\left(\frac{\partial}{\partial
x}-\frac{1}{x}\right), \quad  L_n=\quad x^{2n}\nabla\frac{1}{x^{2n}}\nabla
\label{A1}
\end{equation}
The intertwiner can be used to produce eigenfunctions of the interlaced operators from eigenfunctions of Laplacian:
$$
\Delta \phi = \lambda \phi  \qquad \Longrightarrow L T[\phi]=\lambda T[\phi]
$$
The classification of QAIS is an open problem. Presently, the algebraic integrability is found in systems related with finite-reflection Coxeter groups and their special deformations \cite{OP, C, CFV, CFV1}. In two dimensions, the most general class of QIAS is related to soliton solutions of the KDV hierarchy \cite{B, BL}, while the algebraically integrable systems with coefficients depending on one variable only are related to rational solutions of the KDV hierarchy (the Adler-Moser polynomials) \cite{BC, AM}.

In the following section we review the applications of QAIS to generalization of
harmonic analysis emerging in the free-boundary problems.

\end{section}

\begin{section}{  Mean value integral identities, conservation laws\\ and
algebraic domains}\label{quadrature}

Let $\phi(z,\bar z)$ be a time-independent function satisfying
\begin{equation}
\nabla\kappa\eta\nabla\phi=0, \quad z\in\Omega \label{elliptic}
\end{equation}
in the whole $\Omega$, including $z=z_1$.

Let us now estimate the time derivatives of the following quantities
$$
M[\phi]=\int_{\Omega(t)}\eta\phi dxdy.
$$
By considering an infinitesimal variation of the fluid domain
$\Omega(t)\to \Omega(t+dt)$, we get
$$
\frac{d M[\phi]}{dt}=\oint_{\partial \Omega(t)}v_n\eta\phi dl,
$$
where $dl$ is the boundary arc length. From (\ref{darcy}),
(\ref{boundary}), (\ref{kinematik}) it follows

$$
\frac{d M[\phi]}{dt}=\oint_{\partial
\Omega(t)}\left(\phi\kappa\eta\nabla
P-P\kappa\eta\nabla\phi\right)\cdot ndl.
$$
By applying the Stokes theorem and recalling that $P$ and $\phi$
satisfy (\ref{Pequation}), (\ref{elliptic}), we get
\begin{equation}
\frac{dM[\phi]}{dt}=2\pi \hat q^*[\phi](z_1,\bar z_1)
\label{dynamics}.
\end{equation}
Note that mixed derivatives are absent in $\hat q$ (c.f. (\ref{q})),
for by (\ref{elliptic}), $\frac{\partial^2\phi}{\partial z\partial
\bar z}$ is expressed through first derivatives of $\phi$. $\bar
q_j$ is the complex conjugate of $q_j$, since both $\phi(z,\bar z)$
and $\bar\phi(\bar z,z)$ satisfy (\ref{elliptic}).

It follows that $M[\phi]$ is conserved for any solution of
(\ref{elliptic}), a such that $\hat q^*[\phi](z_1,\bar z_1)=0$.

The conservation laws have been first obtained for the homogeneous
medium flows in \cite{R}, the variable-coefficient generalization
seems to be first presented in \cite{EE}.

The Laplacian growth (\ref{homogenous})
is the simplest example where the conservation laws can be written
down explicitly \cite{R}. In this example, any (anti)analytic in
$\Omega$ function satisfies (\ref{elliptic})
\begin{equation}
\phi(z, \bar z)=f(z)+g(\bar z), \quad {\rm for} \quad \kappa=1,
\quad \eta=1 \label{fu},
\end{equation}
where $f,g$ are (anti)analytic in $\Omega$ and the quantities
$$
\int_{\Omega(t)}\left(f(z)(z-z_1)^{k+1}+g(\bar z)(\bar z-\bar
z_1)^{k+1}\right)dxdy
$$
are integrals of motion for the free-boundary flows driven by a
multipole source of order $k$ located at $z=z_1$ in the homogeneous
medium.

Returning to the general case, we integrate (\ref{dynamics}) and get
$$
M[\phi](t)=M[\phi](0)+2\pi \hat Q[\phi](z_1,\bar z_1),
$$
where
\begin{equation}
\hat Q=\int_{0}^{t} \hat q^*(t')dt'= Q+\sum_{j=1}^k\left(
Q_j\frac{\partial^j}{\partial z^j}+\bar
Q_j\frac{\partial^j}{\partial \bar z^j}\right), \quad
\label{integrals}.
\end{equation}
Therefore $M[\phi](t)$, and consequently the shape of the domain, does
not depend on the history of sources and is a function of
``multipole fluxes"
$$
Q=\int_0^t q(t')dt', \quad Q_j=\int_{0}^tq_j(t')dt', \quad \bar
Q_j=\int_{0}^t\bar q_j(t')dt', \quad j=1..k
$$
injected by the time $t$.

Now consider the special case when $M[\phi](0)=0$. This describes the
injection of the fluid to an initially empty medium. In such a case we have
\begin{equation}
\int_\Omega \eta(z,\bar z)\phi(z,\bar z)dxdy=2\pi \hat
Q[\phi](z_1,\bar z_1), \label{simple}
\end{equation}
Equation (\ref{simple}) is a generalization of the mean value or quadrature identities
appearing in the theory of harmonic functions \cite{S} to the case of elliptic equations with
variable coefficients. The simplest example of a quadrature identity is a mean value theorem for harmonic functions. Special domains for which the quadrature
identities hold are called quadrature domains in the theory of the
harmonic functions.

Note, that the above derivation can be also adapted to the higher dimensional case where similar identities hold.

Now, we return to our inverse free-boundary problem stated in the introduction. It can be rewritten in the language of quadrature domains as follows: Given a domain approximated by polynomial conformal mapping (\ref{conformal}), find $\kappa$ and $\eta$ that best approximate given permeability and porosity and for which a quadrature identity holds in the above domain.

Below, we are going to consider the case of stratified media, when the permeability and porosity depend on one coordinate only.

\end{section}

\begin{section}
{ Planar free-boundary flows in stratified media and rational
solutions of the KDV hierarchy}

Below we consider the solution to the inverse problem for a stratified medium described in the introduction.

\begin{theorem}

Let $\Omega(t)$ be a domain in the plane given by a polynomial conformal mapping of the unit disc (\ref{conformal}).
Then this domain can be formed by injection of fluxes $Q$ and $Q_j, j=1..k$ through a multipole source of order $k$ at $z=z_1$ in a medium with permeability and porosity given by (\ref{etakappa}). The order of the source is given by (\ref{sources}).

\end{theorem}

The proof is based on a derivation of a linear system of equations for values of fluxes $Q, Q_j$ that is a verification of quadrature identity (\ref{simple}) for all regular in $\Omega$ solutions of the elliptic equation
\begin{equation}
\xi(x)^2\nabla\frac{1}{\xi(x)^2}\nabla\phi=0
\label{xelliptic}
\end{equation}
with $\xi$ given by (\ref{etakappa}) in an algebraic domain defined by mapping (\ref{conformal}). The "$x$-dependent" part of the above elliptic operator $\xi(x)^2\partial_x\xi(x)^{-2}\partial_x$ is gauge equivalent to a Schrodinger (Lax) operator connected with rational solutions of the KDV hierarchy with potential expressed through Adler-Moser polynomials \cite{AM}.

Our proof follows general argumentation that was used for systems
with finite reflection invariance in \cite{L}. Afterwards we outline a
procedure needed for finding values of fluxes.

In the case of equation (\ref{xelliptic}), the intertwiner between the elliptic operator in (\ref{xelliptic}) and Laplacian (cf (\ref{ii})) can be expressed in the form of Wronskian \cite{AM, BC}
\begin{equation}
T[f](x,y)=\frac{W[\psi_1(x),\dots,\psi_n(x), f(x,y)]}{p_{n-1}(x)}
\label{W}
\end{equation}
where the Adler-Moser polynomials $p_n$ are also expressed as Wronskians
$$
p_n(x)=W[\psi_1(x),\dots,\psi_n(x)]
$$
of functions defined by
$$
\psi_{i+1}(x)^{''}=\psi_i(x), \quad \psi_1(x)=x
$$
The $n$th Adler-Moser polynomial depends non-trivially on $n-1$ free parameters (so called "KDV times") that emerge as integration constants for $\psi_i$th in the above equation.

It then follows that the general solution of (\ref{xelliptic}) can be written down as
$$
\phi(x,y)=T[f(z)+g(\bar z)]
$$
with $T$ given by (\ref{W}) and $f(z), g(\bar z)$ being functions (anti)analytic in $\Omega$. Taking (\ref{etakappa}) into account we get
$$
\eta\phi=S(x)W[\psi_1(x),\dots,\psi_n(x), f(z)+g(\bar z)]
$$
which is a linear differential operator with polynomial in $z, \bar z$ coefficients  acting on $f(z)$ or $g(\bar z)$. It is sufficient to consider its action on $f(z)$:
$$
\eta\phi=\sum_{i=0}^n P_j(z, \bar z)\frac{d^jf(z)}{dz^j}.
$$
We can now evaluate lhs of the quadrature identity (\ref{simple}). By using the Green's theorem, we get
\begin{equation}
\int_\Omega \eta\phi dxdy=\sum_{j=0}^n\oint_{\partial\Omega} R_j(z, \bar z)\frac{d^jf(z)}{dz^j} dz, \quad \frac{\partial R_j(z,\bar z)}{\partial \bar z}=\frac{1}{2i}P_j(z, \bar z)
\label{quadl}
\end{equation}
where
\begin{equation}
R_j=c_j\bar z^{1+\deg(S)+j(j+1)/2} + {\rm lower} \,\, {\rm order} \,\, {\rm terms} \,\, {\rm in} \quad \bar z, z
\label{polynoj}
\end{equation}
is a polynomial of $\deg(S)+1+j(j+1)/2$ th power in $\bar z, z$.

Since $\partial\Omega$ is an image of unit circle under the mapping (\ref{conformal}), and since on the unit circle
$$
\bar w=1/w, \quad \bar{F(w)}={\bar F} (1/w)
$$
we can rewrite (\ref{quadl}) as integrals taken around the unit circle
$$
\int_\Omega \eta\phi dxdy=\sum_{j=0}^n\oint_{|w|=1}  F^\prime(w) R_j(F(w), \bar F(1/w))\left(\frac{1}{F^\prime(w)}\partial_w\right)^j[f(F(w))]dw.
$$
The map (\ref{conformal}) is analytic and polynomial with $F^\prime(w)\not=0$ on the unit disc, and since $R_j(z,\bar z)$ are polynomials in $z, \bar z$, the above integrals can be evaluated as a sum of a finite number of residues at $w=0$. By taking (\ref{polynoj}) and (\ref{conformal}) into account, we get
$$
\int_\Omega \eta\phi dxdy=\sum_{j=0}^{(\deg(S)+1+n(n+1)/2)\deg F+n-1} C_j\left(\frac{d^jf(F(w))}{dw^j}\right)_{w=0}
$$
Now we consider rhs of the quadrature identity (\ref{simple}).
According to (\ref{integrals}) we have
$$
\hat Q[\phi](z_1,\bar z_1)=\hat Q [T[f(z)]]_{z=z_1}=
$$
$$
Q\sum_{j=1}^n A_j\left(\frac{d^jf(F(w))}{dw^j}\right)_{w=0}+\sum_{j=1}^k \sum_{m=0}^n(Q_jB_{jm}+\bar Q_jC_{jm}) \left(\frac{d^{j+m}f(F(w))}{dw^{j+m}}\right)_{w=0}.
$$
Equating the above expressions for lhs and rhs of the quadrature identity, we get $k+n+1$ equations for $2k+1$ unknowns $Q, Q_j, \bar Q_j, j=1..k$ with $k$ given by (\ref{sources}). The condition that $\bar Q_j$ is a complex conjugate of $Q_j$ and $Q$ is real uniquely defines all fluxes.

In concluding the consideration of highly viscous planar flows, it is
interesting to note that the Adler-Moser polynomials, and the rational
Baker-Akhieser function of the KDV hierarchy in general, seem to be
ubiquitous in integrable problems of two dimensional fluid
mechanics: They also emerged on the other extreme, i.e. in the theory of
inviscid flows that are finite-dimensional reductions of the Euler
equations, where they describe equilibria and uniform motion of vortices in
two dimensions \cite{Ba, L1, L2}.

\end{section}

\begin{section}
{ Higher-Dimensional Flows  }

Three-dimensional viscous free-boundary flows are also of practical
interest, while in dimensions four and higher the interest is mainly
due to generalization of theory of harmonic functions and quadrature
identities.

Since the conformal mapping technique is unavailable here, we
propose to write down the multidimensional quadrature identities using
methods of construction of fundamental solutions for QAIS \cite{BM}.

Let us demonstrate the application of these methods by the example of an expanding sphere
$(\zeta_1-\zeta_1^\prime)^2+(\zeta_2-\zeta_2^\prime)^2+\dots+(\zeta_d-\zeta_d^\prime)^2=r(t)^2$
centered at $\zeta^\prime=(\zeta_1^\prime, \dots, \zeta_d^\prime)$
in $d$-dimensional Euclidian space. The flow is governed by the
following elliptic equation
\begin{equation}
{\cal L}[P](\zeta)=\sigma_{d-1}\hat q[\delta(\zeta_1-\zeta_1^\prime)\dots \delta(\zeta_d-\zeta_d^\prime)],
\label{elliptic_multi}
\end{equation}
where $\sigma_{d-1}$ is the area of $d-1$ dimensional unit sphere
and
$$
{\cal L}=\nabla \frac{1}{\zeta_1^2} \nabla, \quad \eta=1, \quad \nabla:=\left(\partial/\partial \zeta_1, \dots \partial/\partial \zeta_d\right).
$$
This is the simplest non-trivial example of the flow in non-homogeneous medium in higher dimensions that is related to QAIS. Similar results can be obtained for more complex cases of QAIS, such as those related to rational or soliton solutions of the KDV hierarchy or finite-reflection groups.

\begin{proposition}
Let $\phi$ be any solution of $\nabla \zeta_1^{-2} \nabla\phi=0$ regular in the closure of the ball
$$
\Omega:\, (\zeta_1-\zeta_1^\prime)^2+(\zeta_2-\zeta_2^\prime)^2+\dots+(\zeta_d-\zeta_d^\prime)^2<r^2.
$$
Then
$$
\int_\Omega \phi(\zeta) d \zeta_1 \dots d \zeta_d = {\it v_d} \Bigg( \phi(\zeta) + \frac{r^2}{(d+2)\zeta_1^\prime} \frac {\partial
\phi(\zeta)}{\partial \zeta_1} \Bigg)_{\zeta=\zeta^\prime},
$$
where $v_d$ is volume of $\Omega$, and
$$
\int_{\partial\Omega} \phi(\zeta) d s = {\it s_{d-1}} \Bigg(
\phi(\zeta) + \frac{r^2}{\zeta_1^\prime d} \frac {\partial
\phi(\zeta)}{\partial \zeta_1} \Bigg)_{\zeta=\zeta^\prime},
$$
where integration  is taken over the surface of $(d-1)$ dimensional sphere
$\partial\Omega$ centered at $\zeta=\zeta^\prime$, and $s_{d-1}$
denotes the area of this sphere.
\end{proposition}

This is an analog of the mean value theorem for harmonic functions.

Proof: is by direct solution of the free-boundary problem (\ref{elliptic_multi}), (\ref{boundary}) in $d$ dimensions. Below we present solution for all dimensions except for $d=2$ and $d=4$. The proofs for the above dimensions differ by technical details only.

Let us construct a solution assuming that operator $\hat q$ in (\ref{elliptic_multi}) is of the first order. We may write down solution of (\ref{elliptic_multi}) in the following form,
\begin{equation}
P(\zeta)=\alpha G(\zeta, \zeta^\prime)+\beta \frac{\partial G(\zeta, \zeta^\prime)}{\partial \zeta_1^\prime} + \Psi(\zeta; \zeta^\prime)
\label{sum}
\end{equation}
where $\alpha$ and $\beta$ are $\zeta$-independent parameters, $\Psi(\zeta; \zeta^\prime)$ is a homogenous solution, i.e. ${\cal L}[\Psi(\zeta; \zeta^\prime)]=0$, and $G$ is a fundamental solution
\begin{equation}
{\cal L}[G(\zeta,\zeta^\prime)]=\sigma_{d-1}\delta(\zeta_1-\zeta_1^\prime)\dots \delta(\zeta_d-\zeta_d^\prime).
\label{solution_L}
\end{equation}
Let us also introduce the fundamental solution of the Laplace
operator in $d$-dimensions
\begin{equation}
\Delta[\mathcal{G}_0(\zeta,\zeta^\prime)]=\sigma_{d-1}\delta(\zeta_1-\zeta_1^\prime)
\dots \delta(\zeta_d - \zeta_d^\prime), \quad
\mathcal{G}_0=\rho^{2-d}, \quad
\rho=\sqrt{(\zeta_1-\zeta_1^\prime)^2+\dots+(\zeta_d-\zeta_d^\prime)^2}
\label{coulomb}
\end{equation}
The fundamental solution $G$ of ${\cal L}$ can be found applying the Hadamard's expansion, the corresponding procedure was developed for QAIS in \cite{BM, B1}. Here we briefly describe it:

Consider a QAIS with the Hamiltonian
\begin{equation}
H=\Delta+2\Delta[\log \tau(\zeta)]=\xi(\zeta)\nabla\xi(\zeta)^{-2}\nabla\xi(\zeta).
\label{H}
\end{equation}
The corresponding free-boundary problem is governed by the elliptic operator
\begin{equation}
{\cal L}=\nabla\xi(\zeta)^{-2}\nabla,
\label{QAIS}
\end{equation}
where $\tau(\zeta)$ is a polynomial in $\zeta_1, \dots, \zeta_d$.
According to \cite{BM, B1}, see Appendix, the fundamental solution
(\ref{solution_L}) can be written explicitly as follows
\begin{equation}
G(\zeta,
\zeta^\prime)=\frac{\xi(\zeta)\xi(\zeta')}{\tau(\zeta')}\sum_{i=0}^n
{\cal T}_i\Bigg[\Delta^{-i}[\mathcal{G}_0(\zeta,\zeta^\prime)]\Bigg]
, \label{Hadamards}
\end{equation}
where $n=\deg(\tau(\zeta))$, and operators ${\cal T}_i$, acting on
$\Delta^{-i}[\mathcal{G}_0]$, are defined recursively as
\begin{equation}
{\cal T}_{i+1}={\cal T}_i\Delta-H{\cal T}_i, \quad {\cal T}_0=\tau(\zeta).
\label{Ti}
\end{equation}
When ${\cal T}_i$ vanish for $i>n$, the Hadamard's expansion (\ref{Hadamards}) truncates, which is a property of QAIS. Note that, as follows from the last equation, ${\cal T}_n$ is nothing but an intertwinning operator and one can also consider recursive procedure (\ref{Ti}) as a test for algebraic integrability of a system with given $\tau(\zeta)$ (or $\xi(\zeta)$) accompanied by simultaneous construction of the intertwinning identity $H{\cal T}_n={\cal T}_n\Delta$.

In our case $\tau(\zeta)=\xi(\zeta)=\zeta_1$, $n=1$, and
$$
{\cal T}_0=\zeta_1, \quad {\cal T}_1=2\frac{\partial}{\partial \zeta_1}-\frac{2}{\zeta_1}.
$$
Taking into account that in $d$ dimensions (except $d=2$ and $d=4$)
$$
\Delta^{-1}[\mathcal{G}_0](\rho)=\frac{-\rho^2}{2(d-4)}\mathcal{G}_0(\rho),
$$
we get
\begin{equation}
G=\left(\zeta_1\zeta_1^\prime-\frac{\rho^2}{d-4}\right)\rho^{2-d} .
\label{G}
\end{equation}
By substituting
\begin{equation}
\alpha=1, \quad \beta=\frac{r^2}{\zeta_1^\prime d}, \quad \Psi(\zeta;\zeta^\prime)=-\frac{\rho^2+2\zeta_1\zeta_1^\prime+(d-2)\zeta_1^2}{r^{2-d}d}+\frac{r^{4-d}}{d-4}
\label{coefficients}
\end{equation}
into (\ref{sum}), and taking (\ref{G}) into account, we get
$$
P(\zeta)=\left(\zeta_1\zeta_1^\prime-\frac{\rho^2}{d-4}\right)\rho^{2-d}-\frac{r^2}{d}\Bigg((d-2)\zeta_1(\zeta_1^\prime-\zeta_1)\rho^{-d}-\rho^{2-d}\Bigg)
-\frac{\rho^2+2\zeta_1\zeta_1^\prime+(d-2)\zeta_1^2}{r^{2-d}d}+\frac{r^{4-d}}{d-4}
$$
It follows that $P$ vanishes at the boundary $\rho=|\zeta-\zeta^\prime|=r$ and therefore $P$ is a solution of the free-boundary problem for expanding sphere. From (\ref{sum}), (\ref{solution_L}) we get
$$
{\cal L}[P](\zeta)=\sigma_{d-1}\left(1-\frac{r^2}{\zeta_1^\prime d}\frac{\partial}{\partial\zeta_1}\right)[\delta(\zeta_1-\zeta_1^\prime) \dots \delta(\zeta_d - \zeta_d^\prime)].
$$
By applying the procedure of derivation of quadrature identities, given in section \ref{quadrature}, and taking the above equation into account, we complete the proof of the Proposition.

Similar propositions could be proved for other $d$-dimensional analogs of planar integrable elliptic growth systems mentioned in this paper.

In the planar case, rather than using conformal mappings, one can define the polynomial domains as the Laplacian quadrature domains formed by fluid injected through a multipole source into an initially empty homogeneous medium. Such a definition of the polynomial domains is acceptable in any dimension.

It is, however, unclear whether the higher-dimensional analogs of theorems considered in this paper are also valid for other than spherical Laplacian quadrature domains in $d>2$.

In general, the proofs use the following arguments: By the Green's theorem
$$
\int_\Omega \left(P(\zeta){\cal L} [G](\zeta, \tilde \zeta)-G(\zeta, \tilde \zeta){\cal L}[P](\zeta)\right)d^d\zeta=\int_{\partial\Omega}\left(P(\zeta)\xi(\zeta)^{-2}\partial_n G(\zeta, \tilde \zeta)-G(\zeta, \tilde \zeta)\xi(\zeta)^{-2}\partial_n P(\zeta)\right)ds
$$
and by (\ref{darcy}), (\ref{boundary}), (\ref{elliptic_multi}), (\ref{solution_L}) we have
$$
P(\zeta)=\hat q[G](\zeta, \zeta')-\int_{\partial\Omega}G(\zeta, \tilde \zeta)\eta(\tilde \zeta)v_n(\tilde \zeta)d\tilde s.
$$
Taking again into account the boundary condition (\ref{boundary}), we get the equation for (constant) coefficients of $\hat q$:
\begin{equation}
\hat q[G](\zeta, \zeta')=\int_{\partial\Omega}G(\zeta, \tilde \zeta)\eta(\tilde \zeta)v_n(\tilde \zeta)d\tilde s, \quad \zeta, \tilde\zeta \in\partial\Omega .
\label{multi_solution}
\end{equation}
To further simplify solution of this system one has to use (\ref{Hadamards}) and the properties of Laplacian domains (e.g. $v_n=dr(t)/dt$ in the case of the sphere).

\end{section}

\begin{section}{Conclusion and open questions}

In this article we have considered new applications of the theory of Quantum Algebraically Integrable Systems (QAIS) to the fluid dynamics of free boundary viscous flows. In two dimensions, we used conformal mapping techniques in combination with conservation laws in the form of quadrature identities to solve whole classes of inverse free-boundary problems. In higher-dimensions we applied a "brute force" approach to solve free-boundary problems directly. This approach relies on methods developed in the theory of QAIS, which limits us to a quite restricted class of problems. It will be interesting to find more general and perhaps less technical approaches for the higher-dimensional cases.

It is worthy to mention the related classification problems, namely the problem of description of all elliptic operators (or media, in the language of fluid dynamics) for which the polynomial dynamics of free boundary (\ref{conformal}) can be sustained by a finite number of sources. This question seems to be connected with an open problem of classification of QAIS \cite{BL, CFV, CFV1}.

\end{section}

\vspace{8mm}

{\bf\large Appendix. Fundamental solutions  and test for algebraic
integrability}

\vspace{5mm}

The basis of the  method of constructing fundamental solutions for QAIS
\cite{BM}  is the use of an intertwining operator with spectral
parameter $\lambda$. Let us consider the identity
\begin{equation}
- e^{-H \frac{\partial}{\partial \lambda} } \lambda \theta(\zeta,
\zeta') e^{\Delta \frac{\partial}{\partial \lambda}} = (H-\lambda)D
= D (\Delta -\lambda),
\label{interD}
\end{equation}
where
\begin{equation} D=e^{-H \frac{\partial}{\partial \lambda} }
\theta(\zeta, \zeta') e^{\Delta \frac{\partial}{\partial \lambda}}
\label{D}
\end{equation}
In case of QAIS, one can find such $\theta(\zeta, \zeta')$ that intertwining operator $D$
has a finite order in $\frac{\partial}{\partial \lambda}$ and leaves
invariant the $\delta$-function concentrated at the point
$\zeta=\zeta'$:
\begin{equation}
D  \delta(\zeta-\zeta') = \delta(\zeta-\zeta') \label{Ddelta}
\end{equation}
The fundamental solution of $H-\lambda$ can be therefore obtained by action of $D$ on the fundamental solution $\mathcal{G}_{\lambda}(\zeta,\zeta')$ of
operator $\Delta-\lambda$:
\begin{equation}
\mathcal{G}_{\lambda}= \sum_{i=0}^{\infty} \lambda^i \Delta^{-i}
[\mathcal{G}_0], \label{Gl}
\end{equation}
where $\Delta^{-i} [\mathcal{G}_0]$ denotes the fundamental solution of
$\Delta^{i+1}$, i.e. $\Delta [\Delta^{-i} [\mathcal{G}_0]]=
\Delta^{-i+1} [\mathcal{G}_0]$.

The $\partial_\lambda$-expansion of the intertwining operator (\ref{D})  is
\begin{equation}
D=\sum_{i=0}^{\infty} \frac{1}{i_{ } !} \mathcal{T}_i
\frac{\partial^i}{\partial \lambda^i},
\label{Dl}
\end{equation}
where operators $\mathcal{T}_i$ can be defined recursively as
$$
\mathcal{T}_{i+1}= \mathcal{T}_i \Delta -H \mathcal{T}_i, \quad
\mathcal{T}_0=\theta(\zeta, \zeta').
$$
From (\ref{Ddelta}) it follows that at $\zeta=\zeta'$,
$\theta=1$.

The intertwining operator $D$ is the finite order differential
operator in $\lambda$, if the above expansion truncates at $i=n$, i.e. when
$\mathcal{T}_{n+1}=0$ for a finite $n$. In such a case, we have the
intertwining identity
$$
H \mathcal{T}_n = \mathcal{T}_n \Delta.
$$
Since, according to (\ref{interD}), the fundamental solution $\Phi$
of the equation $H\Phi(\zeta, \zeta')= \delta(\zeta-\zeta')$ equals to
$D \mathcal{G}_{\lambda}$ evaluated at $\lambda=0$, then from (\ref{Gl})
and (\ref{Dl}) we get
$$
\Phi(\zeta, \zeta')=\sum_{i=0}^n {\cal
T}_i\Bigg[\Delta^{-i}[\mathcal{G}_0(\zeta,\zeta^\prime)]\Bigg].
$$
In the case, when the truncation occurs for a $\zeta'$-independent
function $\mathcal{P}(\zeta)$ of degree $n$, i.e. when
$$
\mathcal{T}_{n+1}=0, \quad \mathcal{T}_{i+1}= \mathcal{T}_i \Delta -H \mathcal{T}_i, \quad
\mathcal{T}_0=\mathcal{P}(\zeta),
$$
one can choose
$$
\theta(\zeta, \zeta')=\mathcal{P}(\zeta)/\mathcal{P}(\zeta').
$$
From (\ref{H}), (\ref{QAIS}) it follows that the fundamental
solution $G$ of $\mathcal{L}$ is related to that of $H$ as $G(
\zeta, \zeta')=\xi(\zeta)\xi(\zeta')\Phi( \zeta, \zeta')$ and in the
case when $\mathcal{P}=\tau$ we arrive at equations
(\ref{Hadamards}, \ref{Ti}).

The choice of $\mathcal{P}(\zeta)$ is not unique and is defined modulo
an arbitrary polynomial: When $\mathcal{P}(\zeta)$ is multiplied by
a polynomial of degree $m$ in $\zeta$, the truncation takes place at
$i=n+m$.

Note that the above choice of $\mathcal{P}=\tau$, where $\tau$ is
the potential generating "tau-function" in (\ref{H}), does not always
correspond to the polynomial of minimal degree. For instance, in
the case of QAIS related to finite reflection groups, the degree of
a minimal polynomial can be smaller than that of $\tau$-function. On
the other hand, for QAIS related to generic Adler-Moser polynomials
(generic values of "times" $t_i$), the $\tau$-function is the
minimal polynomial for which truncation occurs.

As an example one can consider system (\ref{A1}) which corresponds
to  degeneration of the Adler-Moser polynomials (with all $t_i=0$),
where the minimal polynomial equals $x^n$, while $\tau=x^{n(n+1)/2}$.
When $t_i\not=0$, the degeneration is lifted and the minimal
polynomial equals an Adler-Moser polynomial without multiple roots
$\tau=p_n(x)=x^{n(n+1)/2}+\dots $.

The condition for a truncation can be considered as a test for
algebraic integrability: It has been shown that the systems
(\ref{H}) for which truncation takes place are algebraically
integrable. Moreover, based on examples of QAIS found so far,  it
has been conjectured, that the converse is also true,
\cite{B}-\cite{CFV}.

Also, the truncation condition turns out to be equivalent to
existence of polynomial solutions of the following "harmonic" chain
of bilinear PDEs \cite{B2} for all non-negative $n$:
$$
\tau_{n} \Delta \tau_{n+1} - 2 ( \nabla \tau_{n} \cdot \nabla
\tau_{n+1}) + \tau_{n+1} \Delta \tau_{n} = 0, \quad \tau_0=1.
$$
For instance, in the case of one dimension this bilinear equation
becomes a recurrent relation for the Adler-Moser polynomials (\ref{BC}),
which has the only chain of consecutive solutions. In the
multi-dimensional case the chain of solutions branches, giving,
according to the above mentioned conjecture, all $\tau$-functions of
algebraically integrable Schr\"odinger operators (\ref{H}).

Concluding, we give the list of all polynomial $\tau$-functions of
irreducible rational algebraically integrable Shr\"odinger operators
(\ref{H}) \footnote{Irreducible QAIS is a system which is not a
direct sum of QAIS of smaller dimensions} found so far:

\begin{itemize}

\item One-dimensional QAIS related to rational solutions of KdV
$$
\tau=p_n(x)
$$

\item Two-dimensional QAIS related to soliton solutions of KdV \cite{BL}
$$
\tau(x,y)=r^{\sum_{i=1}^n k_i} W[\cos(k_1\varphi+\phi_1), \dots , \cos(k_n\varphi+\phi_n)],
$$
where $x=r\cos(\varphi)$, $y=r\sin(\varphi)$, $\phi_1 \dots \phi_n$
are arbitrary initial phases and $0 \le k_1<k_2<\dots<k_n, \quad
k_i\in \mathbb{Z}$.

This class includes a sub-class of Coxeter-invariant (see below) systems in two-dimensions.

\item QAIS in $d$-dimensions ($d>2$) related to the Coxeter root systems or their special deformations
$$
\tau(\zeta)=\prod_{\alpha}(\alpha\cdot\zeta)^{m_\alpha}, \quad \zeta \in \mathbb{C}^d, \quad m_\alpha \in \mathbb{Z}_+,
$$
where product is taken over a finite set of vectors ("positive
roots") $\alpha \in \mathbb{C}^d$. Roots form either a
reflection-invariant Coxeter system (with reflection invariant
multiplicities $m_\alpha$), or special deformations of a Coxeter
root system \cite{CFV1}, \cite{CFV}. In the case of deformations the
roots depend on multiplicities.

\end{itemize}


\end{document}